\documentstyle[12pt,aasms4]{article}

\begin{document}
\title{RAPID VARIABILITY OF GAMMA-RAY BLAZARS:
A MODEL FOR MARKARIAN~421}

\author{M. SALVATI\altaffilmark{1} M. SPADA\altaffilmark{2}
AND F. PACINI\altaffilmark{1,2}}

\altaffiltext{1}{Osservatorio Astrofisico di Arcetri,
                  L. E. Fermi 5, I-50125 Firenze, Italy}
\altaffiltext{2}{Dipartimento di Astronomia e Scienza dello Spazio,
      Universit\`a di Firenze, L. E. Fermi~5, I-50125 Firenze, Italy}

\begin{abstract}
The extremely rapid burst of TeV photons
from Mkn 421 (15 May 1996) can be reconciled
with the standard properties of a relativistic $\gamma$-ray emitting jet
(bulk Lorentz factor $\Gamma \sim10;\ \mbox{size} \sim10^{17}$ cm)
if one assumes that the electrons are accelerated in conical shocks
with both opening and viewing angles $\approx 1/\Gamma$. 
If the injection time and the cooling time are much less than
the photon crossing time, an emission ring moves along the jet and leads to
the appearance of a very rapid flare, in satisfactory agreement with
the observations.
\end{abstract}

\section{INTRODUCTION}

Several blazars, all of them associated with radio loud AGN, show an
intense and variable $\gamma$-ray emission (\cite{zz}).

Various authors (\cite{vv}; \cite{uu}; \cite{tt}; \cite{ss})
have shown that the observed luminosity -- up to
$\sim 10^{48}$\ \rm ergs\ s$^{-1}$ or more -- and variability -- typical time
scales $\sim$ several hours -- can be understood if the emission is due
to inverse Compton scattering of relativistic electrons on softer
photons inside a relativistic jet. These jets move with
bulk Lorentz factor $\Gamma \sim 5-10$, at a distance from the nucleus
of order $10^{16}-10^{17}$ cm:  they appear to be the inner part of the
well known radio jets, with similar $\Gamma$ but larger sizes (10$^{19} -
10^{20}$ cm).  Since the $\gamma$-ray emission occurs closer to the
central objects, it provides direct information about
the physical processes effective in the central region of AGN.

In this connection, the discovery of large (factor of order 20) and
very rapid (time scale around 20 minutes) variations of the TeV emission 
from Mkn~421 on 15 May 1996 (\cite{ee}) represents an important clue to the
physical conditions in the central region of AGN. Variability timescales
of 20 minutes would entail a source size $\approx 10^{15}$ cm
(with problems for photon-photon opacity)
or $\Gamma \stackrel{>}{\sim}10^2$ (with problems for the total energy
and the low probability of observing an event with a visibility cone
$\sim 1/\Gamma$).

Our purpose in this Letter is to present a model for the occurrence of
rapid variations which uses the standard scenario but does not
require extreme values for the Lorentz factor $\Gamma$ or the size.  The
variability is related to anisotropies in the comoving frame of the
jet associated with the geometry of the shocks responsible for the
particle acceleration.  We shall show that  strong spikes in the very
high energy spectrum  are a natural consequence of a nearly conical
shock geometry with characteristic opening $\approx 1/\Gamma$.

\section{RELEVANT TIME SCALES}

We shall assume that  the jet has  a bulk Lorentz factor $\Gamma\gg1$, 
scale length $z$ and radius $r$; the scale length is defined as the
distance over which the bulk properties of the jet change substantially:
let $\Phi$ be one such property, then $z=\vert{\Phi\over d\Phi/dz}\vert$. 
The observer's
line of sight makes an angle $\theta\approx 1/\Gamma$ to the jet axis.

\noindent The time scales relevant to the variability are the following
ones, as measured by a distant observer:

\begin{enumerate}

\item{} the geometrical time needed to the photons to cross the emission
region either in length (longitudinal time $\Delta t_z\approx z/c\Gamma^2$) 
or in width (radial time $\Delta t_r\approx r/c\Gamma$); 

\item{} the radiative time $\Delta t_{em}$ over which the emitting electrons 
lose a large fraction of their energy; if $\gamma$ is the typical electron
Lorentz factor in the comoving frame, the dot indicates the time derivative,
and the prime refers to quantities measured in the comoving frame, then
$\Delta t_{em}\approx(\gamma/\dot\gamma)'/\Gamma$;

\item{} the injection time $\Delta t_{in}$ over which the electrons are fed 
to the emission region; as will be shown in the following, the radiative time 
is typically shorter than the geometrical time, and the electrons must be
accelerated {\it in situ}; there are arguments (\cite{aa}) 
 which point to the electrons being not only accelerated,
but also generated {\it in situ}. 

\end{enumerate}

\noindent From the point of view of the observer, a variability episode appears
convolved with $\Delta t_{in}$, $\Delta t_{em}$, $\Delta t_{r}$, and 
is dominated by the {\it longest} of them. 
On the other hand, in a diverging jet the emission mechanisms become very 
rapidly ineffective over distances longer than $z$, so that $\Delta t_{z}$ 
is an upper bound to the observed variability $\Delta t_{obs}$ 

\begin{equation}
\Delta t_{obs} \approx {\rm min} \bigl[\Delta t_{z}, {\rm max}
(\Delta t_{in}, \Delta t_{em}, \Delta t_{r})\bigr]\ 
\label{eq1}
\end{equation}

\noindent There are various arguments which suggest that $r\approx z/\Gamma$; if
this is indeed the case, and $\Delta t_{r}\approx \Delta t_{z}$ (as we
shall assume in the following), $\Delta t_{obs}$ is necessarily of the
order of the longitudinal time $\Delta t_{z}$. In this paper we show
how a plausible geometric arrangement might release this constraint,
otherwise unavoidable.

The injection time scale $\Delta t_{in}$ is very likely the shortest of
all, of the order of several minutes, depending on the mass one chooses
for the central black hole. The remaining time scales $\Delta t_{em}$
and $\Delta t_z$ are related in an interesting way to the jet optical
depth to photon-photon absorption. This is due to the near equality
of the cross sections involved, and to the radiation density in the jet
being the target for both the emission {\it and} the absorption processes.
The relation can be expressed in closed form under certain assumptions:
only the jet contributes to the radiation field, the radiation spectrum 
is flat in $\nu F_{\nu}$, the scattering process is always in the Thomson
regime. Then one has
\begin{equation}
\ell={{L\sigma_T}\over{mc^3z}},\quad
{{\Delta t_{em}}\over{\Delta t_z}}
={{3\pi}\over{4\ell}}{{\Gamma^3}\over{ \gamma}},\quad
\tau_{\gamma\gamma}={{3\ell}\over{16\pi}}{{\epsilon_u}\over{\Gamma^4}}
\big(\rm{ln}{{\epsilon_u}\over{\epsilon_d}}\big)^{-1}
\label{eq2}
\end{equation}

\noindent The quantity $\ell$ is the compactness of the jet radiation field 
in terms of the \underbar{observed} luminosity $L$,
$\tau_{\gamma\gamma}$ is the jet photon-photon optical depth, and
$\epsilon_u$ and $\epsilon_d$ are the upper and lower energies of the photon
spectrum, respectively, in units of $mc^2$. By multiplying the second and
third of Eqs. \ref{eq2} we finally get

\begin{equation}
{{\Delta t_{em}}\over{\Delta t_z}}\tau_{\gamma\gamma}\approx
{{0.14}\over{ \rm{ln}{{\epsilon_u}\over{\epsilon_d}}}}
{{\epsilon_u}\over{\gamma\Gamma}}
\label{eq3}
\end{equation}

\noindent As long as the right hand side is $\ll 1$, one can have at the same 
time $\tau_{\gamma\gamma}\leq 1$ and a high radiative efficiency
$\Delta t_{em}\ll\Delta t_{z}$. In certain 
BL~Lacs, however, as for instance Mkn~421, the Klein--Nishina regime is
reached and $\epsilon_u/\gamma\Gamma\approx 1$; furthermore, if one requires
that the jet is not loaded with too many pairs (\cite{bb}, \cite{cc}), 
$\tau_{\gamma\gamma}$ has to be $\ll 1$, and a certain
degree of fine tuning becomes necessary. In general, however, $\Delta t_{em}
< \Delta t_z$. The models which have been fitted to typical
$\gamma$-ray blazars have by construction $\Delta t_z$ equal to the 
observed variability time scale, which is several hours; 
Salvati, Spada \& Pacini (1996), for instance, in the case of 3C~279 
find $\Gamma\approx 10$ and $z\approx 10^{17}$~cm.

\section{GEOMETRY AND RAPID VARIABILITY}

The acceleration of the emitting electrons must occur {\it in situ}, and
strong shocks are the most likely location. The orientation of the shocks
with respect to the jet axis is preferentially at angles $\approx1/\Gamma$. 
This is seen in numerical simulations (\cite{ff}, \cite{gg}), 
and is due to arcsin~$1/\Gamma$ being the causal angle traced by a perturbation 
which moves isotropically at speed $c$ in the comoving frame. Also, when
the flow is deflected by the external medium, angles around $1/\Gamma$ are
preferred, since at those angles the ram pressure levels off at values
close to the internal pressure. Finally, it is plausible that $\Delta t_{in}$ 
and $\Delta t_{em}$ are much smaller than $\Delta t_z$.

We can envisage a geometry of the emission region similar to Fig.~\ref{fig1}:
the accelerating shocks are a series of opening and closing cones; a 
very thin layer of extra electrons (the perturbation) move down the jet
with velocity $\beta c,\ \beta=\sqrt{1-1/\Gamma^2}$; they radiate 
immediately after being accelerated, so that the
emission region is a ring at the intersection between the electron layer
and the conical shocks; 
if the cone opening is precisely $\rm arcsin\ 1/\Gamma=\rm arcos \beta$,
and the line of sight coincides precisely with one generatrix of the cone,
the photons emitted in the observer's direction pile up at the same observed 
time.

Many of these restrictions can be released. 
One can assume that the cone opening angle $\alpha$ and the viewing 
angle $\theta$ are different from each other; then the pile up 
condition becomes

\begin{equation}
{{\beta c}\over{{\rm cos}\alpha}}={{c}\over{{\rm cos}(\theta-\alpha)}}\ ,
\qquad \theta=\alpha\pm{\rm arcos}{{{\rm cos}\alpha}\over{\beta}}
\label{eq4}
\end{equation}

\noindent Real solutions exist only if $\alpha\ge{\rm arcos} \beta$;
when $\alpha>{\rm arcos} \beta$, the pile up effect takes place along
{\it two} lines of sight. Also, it is not essential that $\theta$ coincides
exactly with the pile up value. Fig.~\ref{fig2} shows the result of
viewing angles differing by various amounts from the optimal one.
If $x$, $y$, and $z$ are cartesian coordinates of the emission ring,
with $z$ along the jet axis, $s=z~{\rm tan}\alpha$ is the ring radius,
$t$ is the time, and $t_{obs}$ the time at which a given wavefront reaches
the observer, the flux $F$ is given by

\begin{mathletters}
\begin{equation}
F(t_{obs})\propto\int\sqrt{\left({\partial x\over\partial t_{obs}}\right)^2_t 
+ \left({\partial y\over\partial t_{obs}}\right)^2_t }dt
\label{5a}
\end{equation}
\begin{equation}
x={c\over {\rm sin}\theta}[t_{obs}-t(1-\beta {\rm cos}\theta)],\quad
y=\sqrt{s^2-x^2},\quad z=c\beta t
\label{5b}
\end {equation}
\end{mathletters}

\noindent The curves have been computed with $\theta=e\times {\rm arcos}\beta$ 
(where $e$ is a number $0 < e < 1$), $\alpha={\rm arcos} \beta$,
$\Gamma=10$, and $\Delta t_z \approx 9$ hours. One should note
that the vertical axis is logarithmic, and the height of the spikes is
very often a factor of several; one should further note that blazars are
observationally selected to have $\theta\le{\rm arcos} \beta$, and the 
{\it a posteriori} probability of having $\theta\ge e\times{\rm arcos}\beta$
is relatively large, P($\theta\ge e\times
{\rm arcos}\beta~|~\theta\le {\rm arcos}\beta$)~=~$1-e^2$; for instance, the 
sharpest peak shown in Fig.~\ref{fig2} should be observable in 19\% of the cases.
The peaks appear  because the relevant geometry is
not the one of the jet as a whole, but instead the one of the emission
region, which --in the case of strong radiation losses-- traces the surface
of the shocks. The preferential orientation of this surface results in 
the radiation field being strongly anisotropic at any point of the emission
region at the time of the perturbation; the peak photons are distributed
over an angle much narrower than $1/\Gamma$, of the order of $(1/\Gamma)
\times\sqrt{\Delta t_{em}/\Delta t_z}$; the relevant length is the 
radiative length, instead of the geometrical one, and one has
$\Delta t_{obs} \approx \Delta t_{em} = {(z/ c \Gamma^2)}{(\Delta t_{em}
/ \Delta t_z)}$.
Both results can be described in terms of an ``effective'' 
Lorentz factor $\approx\Gamma\times\sqrt{\Delta t_z/\Delta t_{em}}$,
and explain why --at variance with Eq.~ \ref{eq1}-- 
$\Delta t_{obs}\ll \Delta
t_z \approx \Delta t_r$.

Anisotropies in the laboratory frame on scales smaller than $1/\Gamma$
are equivalent to the loss of isotropy in the comoving frame, at variance
with the standard scenario, and may have interesting consequences.
While the peak photons, as we have found, are very closely aligned, the
emitting electrons interact with the same peak photons at angles of the
order of $1/\Gamma$; this breaks the assumptions under which Eq.~\ref{eq3} 
was derived, and allows a situation where {\it at the same time} 
the radiative losses are very rapid and the pair production opacity
is very low.
Also, if the line of sight is not closely aligned with the shock
surface, the observer will infer a photon density at the source {\it lower}
than the one experienced by the electrons; this breaks the assumptions
under which synchrotron self Compton models are usually computed, and allows
larger ratios between the Compton and synchrotron regions of the spectrum.

We stress finally that the sharp features  carry only a small fraction of the
total radiated energy; most of this energy resides in the long
lived, low level tail, which by construction has a time scale of order
$\Delta t_z$. So the observed light curve will exhibit conspicuous sharp 
features only if the injection events are spaced at intervals $\ge \Delta t_z$; 
in the opposite case a given peak will be swamped by the superposition 
of the many tails of the preceding peaks. 

Because of the assumption of negligible radiative lifetime, the 
curves of Fig.~\ref{fig2} are the Green functions of the variability 
problem for the given set of parameters, and by convolving them with an 
arbitrary injection function one obtains the general solution. Fig.~\ref{fig3}
is our best fit by this method to the TeV light curve of Mkn~421 on
15 May 1996 with the
\underbar{same} parameters of Fig.~\ref{fig2} and $e=0.9$. It shows that it 
is not necessary to assume extreme jets to account for extreme variability;
only, similar episodes should not be observed to repeat before times
of order $\Delta t_z\approx$ several hours. We recall that Mkn~421 was
observed to flare with a $\approx 60$ minute timescale also on 7 May 1996
(\cite{ee}). With a different injection function we can easily fit this 
flare as well. Because of the longer duration of the episode, we 
find that the ratio of the fluxes after and before the flare should have been
higher than on 15 May 1996, but there are no data to check this.

\section{CONCLUSIONS}

We have developed a model to explain extremely fast variations in the
emission of relativistic jets. Under the following assumptions:
\begin{enumerate}
\item the jet has typical parameters ($\Gamma\approx 10$, $z\approx
10^{17}$cm, $\Delta t_z\approx 9$ hours);
\item cold electrons are injected in the center and move outwards with
bulk velocity $c\beta$;
\item electrons are accelerated in conical shocks with opening angle
$\approx 1/\Gamma$;
\item injection time and cooling time are $\ll \Delta t_z$;
\item the viewing angle is $\theta\approx 1/\Gamma$,
\end{enumerate}
it is possible 
to observe a very short bump $\ll \Delta t_z$, because the 
component of the peak photons' velocity along the jet is near to the
electron velocity, and the photons pile up on the \underbar{same}
wavefront. In particular, we have been able to fit with typical
jet parameters the 20~minute TeV flare of Mkn~421 on 15 May 1996.

Deviations from
the ideal geometry could be some curvature in the shocks, or a
non-negligible lifetime of the electrons at low energies; these should
result in a certain amount of smoothing, which needs to be computed in
detail.
The extension of this model to the case of non-negligible particle lifetime
is currently under investigation with the aim of comparing 
the expected behaviour of the emission at different frequencies.
We may note straight away that in at least one case (PKS~2155--304)
the burst has been seen to evolve with frequency-dependent time scales
(\cite{qq}),
roughly compatible with synchrotron cooling in a uniform field. This is
possible \underbar{only} if $\Delta t_r\ll \Delta t_z,\ r\ll z/\Gamma$
(with far reaching implications on the jet dynamics), \underbar{or} in a
non-standard geometry as the one proposed here.

\acknowledgements
This work was partly supported by the Italian Space Agency through grants
ASI--95--RS--120 and ARS--96--66.

\newpage
\figcaption[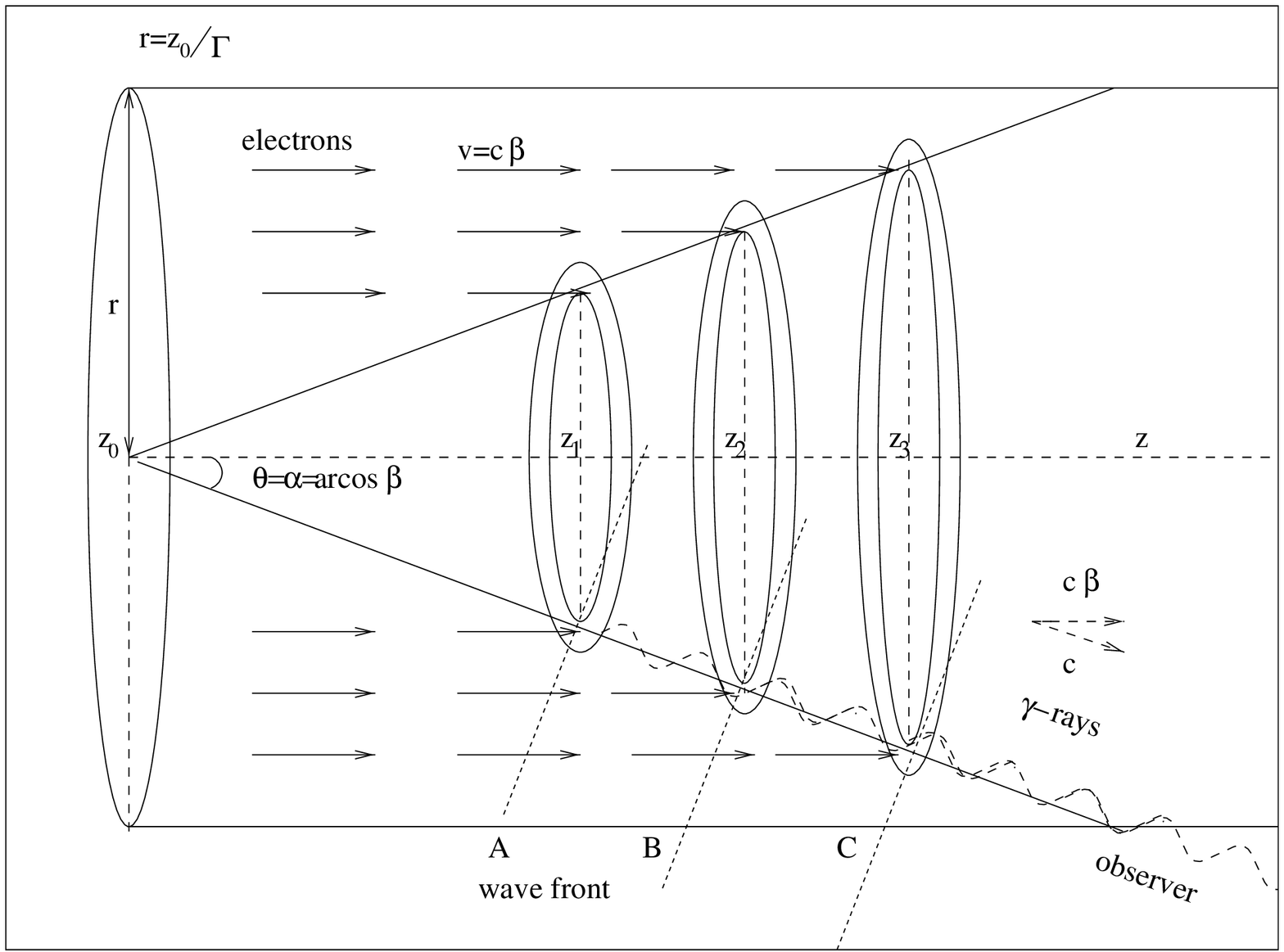]
{The geometry assumed for the acceleration and emission region.
See text for details.\label{fig1}}
\figcaption[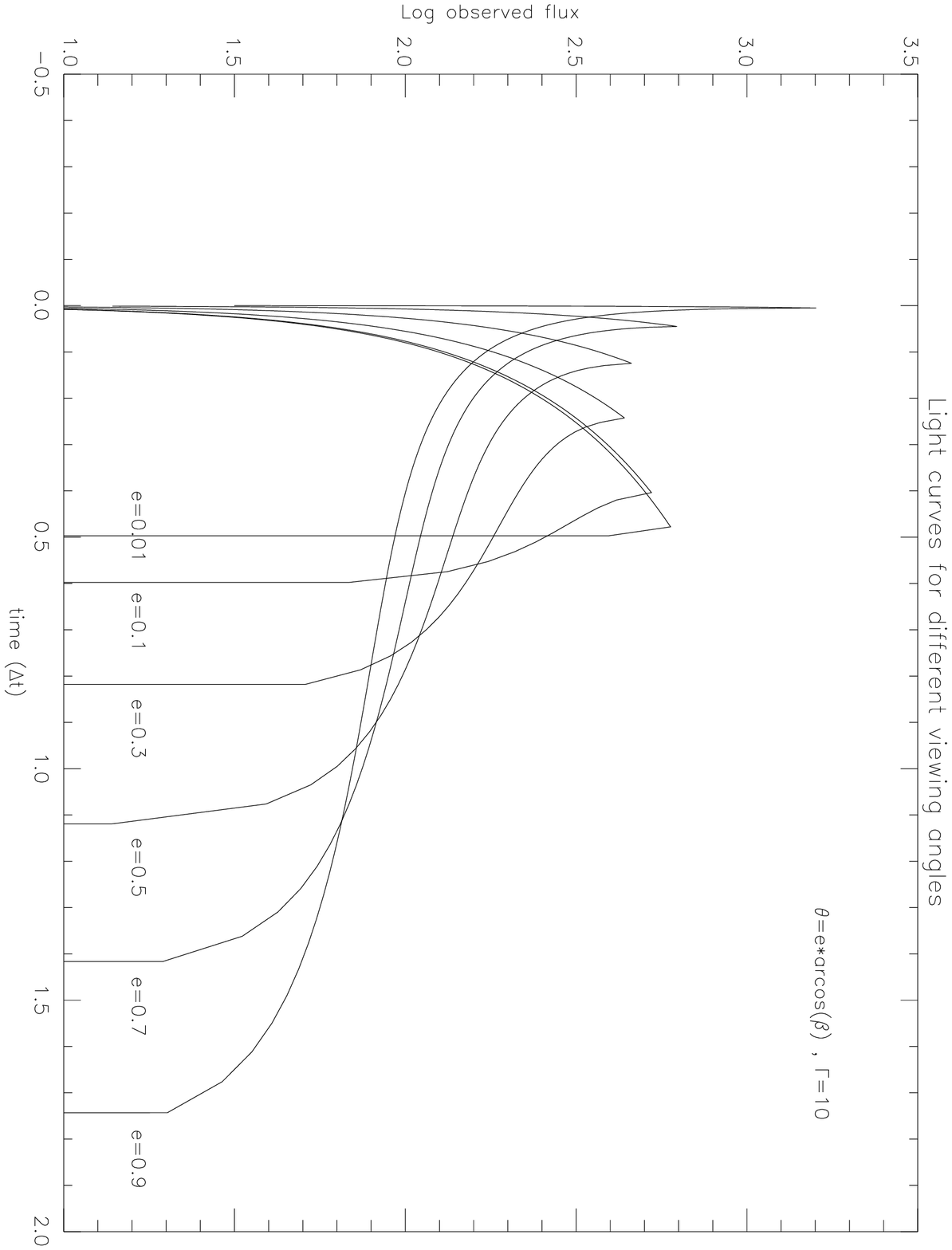]{The light curves observed for $\alpha={\rm arcos}\beta$
and $e= \theta/{\rm arcos}\beta < 1$. The values of $e$ are indicated by
the labels.\label{fig2}}
\figcaption[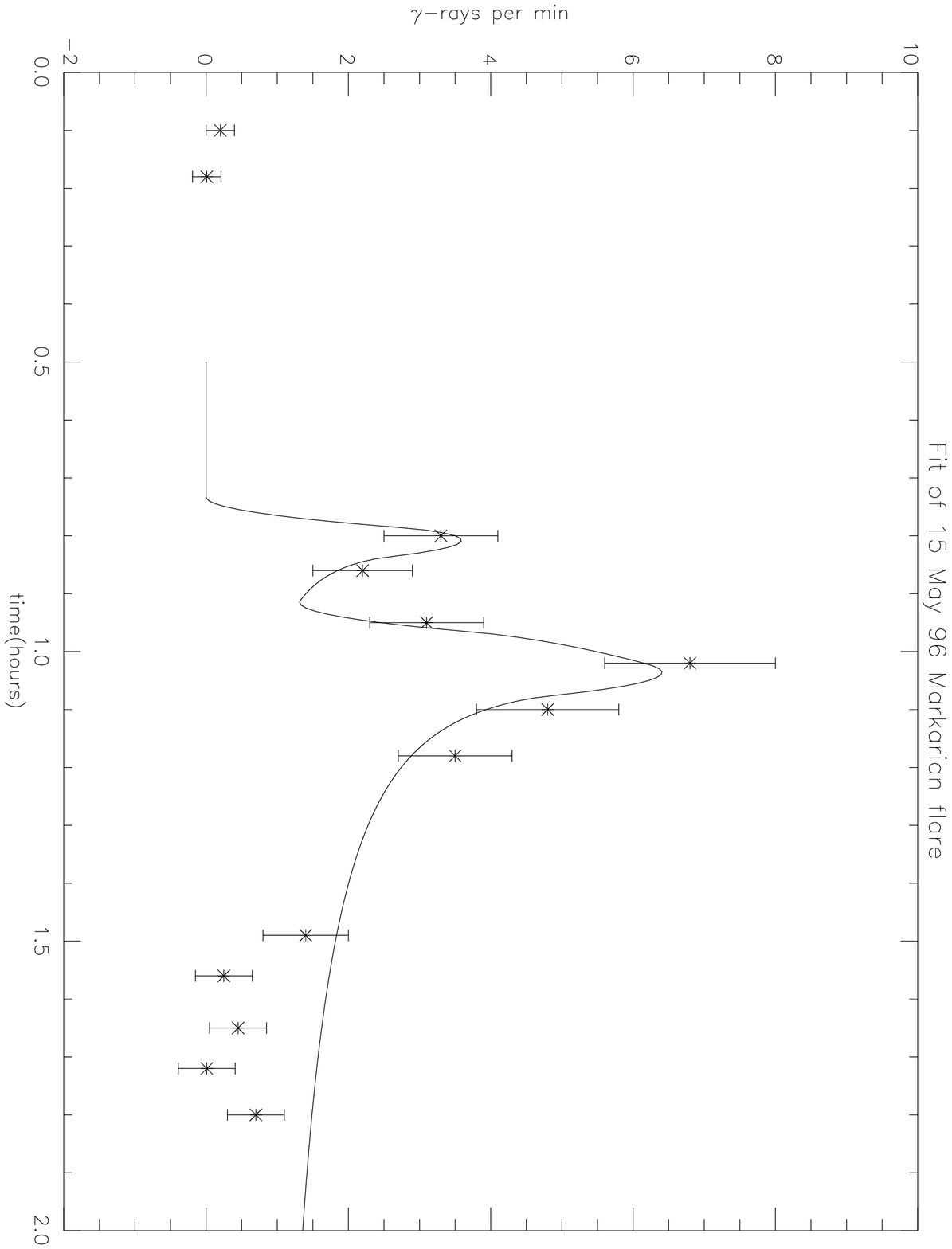]{A fit to the TeV flare of Markarian~421. 
The geometrical time scale of the underlying model is 
$\Delta t_z = 9$ hours.\label{fig3}}

\end{document}